\date{}
\documentclass[12pt]{article}
\usepackage{amsmath,amsthm,amsfonts,amssymb,amsopn,amscd}     
\usepackage{epsfig}
\usepackage{setspace}
\usepackage{stmaryrd}

\DeclareMathOperator{\Tr}{Tr}

\DeclareMathOperator{\supp}{supp}

\def\dim{{\mathrm{dim}}}

\def\Hom{{\mathrm{Hom}}}

\def\id{{\rm id}}

\def\eins{{\mathbf{1}}}
\def\setminus{\smallsetminus}

\def\emptyset{\varnothing}
\def\setminus{\smallsetminus}

\def\Diff{{\mathrm{Diff}}}

\def\RR{{\mathbb R}}
\def\CC{{\mathbb C}}
\def\NN{{\mathbb N}}

\def\ZZ{{\mathbb Z}}

\def\SL{{{\rm SL}}}
\def\PSL{{{\rm PSL}}}
\def\PSU{{{\rm PSU}}}
\def\SU{{{\rm SU}}}
\def\so{{{\rm so}}}
\def\SO{{{\rm SO}}}

\def\HH{{\mathcal H}}

\def\Cf{\mathrm{Conf}}
\def\Pc{P^\uparrow_+}
\def\Lg{{\mathfrak{g}}}
\def\Lh{{\mathfrak{h}}}

\def\sig{\sigma}
 
\def\Psit{\Psi_{\tau\otimes\tau}}\def\Psis{\Psi_{\sig\otimes\sig}}
\def\Ad{\mathrm{Ad}}

\def\eps{\varepsilon} 
\def\inv{^{-1}} 
\def\ol{\overline}
\def\ul{\underline}
\def\wt{\widetilde}

\newcommand{\wick}[1]{{:}#1{:}} 
\def \norm#1{{\vert\!\vert #1\vert\!\vert}}
\def\bpm{\begin{pmatrix}} \def\epm{\end{pmatrix}}

\newcommand{\be}{\begin{equation}} 
\newcommand{\ee}{\end{equation}}
\newcommand{\bea}{\begin{eqnarray}} 
\newcommand{\eea}{\end{eqnarray}}
\newcommand{\ba}{\begin{array}} 
\newcommand{\ea}{\end{array}}
\newcommand{\eref}[1]{Eq.\ \!(\ref{#1})}
\newcommand{\sref}[1]{Sect.\ \!\ref{#1}}

\begin{document}

\title{\vskip-15mm
\LARGE Algebraic conformal quantum field theory \\ 
in perspective}
 
\author{
{\sc Karl-Henning Rehren} \\
\footnotesize Institut f\"ur Theoretische Physik, Universit\"at G\"ottingen,
\\[-1.5mm] \footnotesize Friedrich-Hund-Platz 1, D-37077 G\"ottingen, Germany
\\[-1.5mm] \footnotesize {\tt rehren@theorie.physik.uni-goettingen.de}}

\maketitle

\begin{abstract}
Conformal quantum field theory is reviewed in the perspective of
Axiomatic, notably Algebraic QFT. This theory is particularly
developped in two spacetime dimensions, where many rigorous
constructions are possible, as well as some complete
classifications. The structural insights, analytical methods 
and constructive tools are expected to be useful also for
four-dimensional QFT. 
\end{abstract}

\tableofcontents



\section{Introduction}
\setcounter{equation}{0}
\label{s:intro}
We give an overview of the methods and results of conformal quantum field
theory (CFT), accumulated in the last three decades, in the perspective
of axiomatic approaches. In particular, we advocate the point of view
that a CFT is just a relativistic quantum field theory (QFT) which is
invariant under the group of conformal spacetime symmetries. Thus,
there are no independent ``CFT axioms'', but the usual QFT axioms
apply with an enlarged symmetry.  

Starting from the Wightman axiomatic setting in \sref{s:cft-qft}, we
emphasize the crucial importance of inequivalent representations
(superselection sectors) for several aspects of QFT. This motivates
the formulation in the Haag-Kastler axiomatic setting (``algebraic quantum
field theory'', AQFT), which is particularly powerful to address
superselection sectors. We give a brief review of this setting in
\sref{s:aqft}, and turn to its specific application to chiral CFT in
\sref{s:acft}.  

The sections \sref{s:chimod} and \sref{s:algebraic} give an overview
of various constructive methods to produce models of CFT (and QFT),
and related classification results.

\section{CFT in the context of relativistic QFT}
\setcounter{equation}{0}
\label{s:cft-qft}

Since the conformal group contains the proper orthochronous Poincar\'e
group, a conformal QFT is in particular a relativistic QFT in the usual sense.
The simplest examples are the massless Klein-Gordon and Dirac fields
and the Maxwell field in four spacetime dimensions (4D). Their
conformal symmetry arises as a consequence of the massless field
equations, and not because it was ``postulated'' as an extra feature. 

The Wightman axioms describe quantum fields $\phi$ as operator-valued
distributions on (a common invariant domain in) a Hilbert space $\HH$, 
subject to the principle of Locality (Einstein Causality = commutation
at spacelike distance). The Hilbert space carries a unitary
representation $U$ of the Poincar\'e group $\Pc= \SO(1,D-1)_0\ltimes
\RR^{1,D-1}$ in $D$ dimensions, which
extends to a representation of the conformal group $\Cf_D$ (see below). 
Conformally covariant fields transform with a transformation law    
\be
U_g\phi(x) U_g^*=\alpha_g(\phi(x)) =
D(g,x)\inv \phi(gx),
\ee 
where the fields may be multiplets, and accordingly $D(g,x)$ is a
suitable matrix-valued cocycle. Finally, the vacuum state is a unique
vector $\Omega\in \HH$ which is 
a zero-energy ground state for the Hamiltonian (the generator of the
subgroup of time-translations) in every Lorentz frame, which implies
that it is invariant under $U_g$ for all $g\in \Cf_D$ (in $D>2$).   

For Poincar\'e and scale transformations $g_\lambda:x\mapsto
\lambda x$ ($\lambda\in\RR_+$), the cocycle $D(g,x)$ is
independent of $x$ and is just a matrix representation of
$(\SO(1,D-1)_0\times\RR_+)\ltimes \RR^{1,D-1}$. For scale transformations, one has $D(g_\lambda)\inv=\lambda^d$ where the parameter $d\geq 0$
is the scaling dimension of the field $\phi$.

\subsection{Conformal symmetry}

We take the conformal group $\Cf_D$ to be the connected component of the
group of transformations of $D$-dimensional Minkowski spacetime that
preserve the Minkowski metric $ds^2=\eta_{\mu\nu} dx^\mu dx^\nu$ up to
a positive factor $\omega^2(x)$ that may depend on $x$. This group is
generated (in $D>2$) by the translations
\be
T_a: \quad x^\mu\mapsto x^\mu+a^\mu\qquad (a\in\RR^{1,D-1})
\ee
and the involutive ``conformal inversion''
\be
I: \quad x^\mu \mapsto \frac {x_\mu}{(x\cdot x)},
\ee
such that $S_b=I\circ T_b\circ I$ are the special conformal transformations 
\be
S_b: \quad x^\mu\mapsto \frac{x^\mu - (x\cdot x)b^\mu}{1-2(x\cdot b)+ (x\cdot
  x)(b\cdot b)}\qquad (b\in\RR^{1,D-1}).
\ee
Proper orthochronous Lorentz transformations $x\mapsto\Lambda x$ and
scale transformations $x\mapsto \lambda x$, as well as the inversion
$I$ are generated by translations and special conformal
transformations. 

The singularity of the conformal inversion and the special conformal
transformations can be dealt with as follows (see, e.g., \cite{LM75,BGL93}). 
One introduces the ``Dirac manifold'' which is the projective null cone 
$\{\xi\in \RR^{2,D}:\xi\cdot\xi=0\}/_{\xi=\lambda\xi,\lambda\neq0}\simeq
(S^1\times S^{D-1})/\ZZ_2$ in an auxiliary space with metric
$(+,-\dots-,+)$. $S^1$ is timelike, $S^{D-1}$ spacelike. 
The conformal group acts perfectly regular as
$\Cf_D\sim \SO(2,D)_0/\ZZ_2$ through the linear action of $\SO(2,D)_0$
on the cone. 

Minkowski spacetime is just a (dense) chart with coordinates 
\be
x^\mu = \frac{\xi^\mu}{\xi^D+\xi^{D+1}} \qquad (\mu=0,\dots D-1)
\ee
of the Dirac manifold, so that the singularity of the inversion and
the special conformal transformations is just a coordinate
effect. The Poincar\'e group is the subgroup of $\Cf_D$ preserving the
auxiliary coordinate $\xi^D+\xi^{D+1}$; its generators are
$m^{\mu\nu}$ and $p^{\mu}=\frac12 (m^{\mu D}+m^{\mu D+1})\in \so(2,D)$
($\mu,\nu=0,\dots D-1$). Other Minkowski charts are obtained by acting
with $\Cf_D$ on the Dirac manifold. 

Denoting $p_0$ the generator of the time
translations in the Lie algebra of $\Cf_D$, and $k_0=Ip_0I$ the
generator of the timelike special conformal transformations, 
$h_{\rm conf}:=\frac12(p_0+k_0)$ is the generator of the compact
subgroup $\sim \SO(2)\subset \Cf_D$ in the ``time'' plane
($0$-$D+1$-plane) of the auxiliary space, in which $I$ is the rotation
by $\pi$.

Thus, quantum fields of a QFT with conformal symmetry must be defined as
operator-valued distributions on the Dirac manifold. Their restriction to
any Minkowski chart are Poincar\'e covariant Wightman fields in the
usual sense.   

The discrete spacetime symmetries (parity $P$ and time reflection $T$) are not
part of the axioms; instead, the CPT theorem applies, stating that
while $P$ and $T$ may not be separate symmetries, there is an
antiunitary operator $\Theta$ preserving the vacuum vector and acting
on the fields like the combination of $P$, $T$, and a charge
conjugation $C$. 

Conformal transformations may take pairs of points which are 
timelike separated in a Minkowski chart to pairs of points at spacelike
separation. In other words: the distinction between ``spacelike'' and
``timelike'' is Poincar\'e invariant, but not conformally invariant. 
As a consequence, causal commutativity of fields at spacelike distance
implies causal commutativity also at timelike distance, and the support of
the commutator is constrained to the null cone. 

This conclusion is avoided if one admits projective
representations of the conformal group, such as they occur, e.g., for
the massless Klein-Gordon field in odd spacetime dimensions $D$ which has
half-integer scaling dimension $d=\frac{D-2}2$. In these
cases, fields are not defined on the Dirac manifold, but on a suitable
covering space thereof \cite{LM75}.

\subsection{Two dimensions}
\label{s:2D}

Conformal QFT in two dimensions is in our focus of interest because it
admits a multitude of models which can be rigorously constructed. Most
of these models possess (finitely or infinitely) many positive-energy
representations, which admit, e.g., case studies of the general theory
of superselection sectors as originally formulated for Poincar\'e
covariant QFT in four dimensions (generalizing the univalence
superselection rule or the electromagnetic charge superselection
rule). This theory in particular defines
the statistics (a representation of the permutation group) as an
intrinsic feature of a positive-energy representation. In two
dimensions, a new feature occurs, due to the disconnectedness of the
causal complement of a point in two-dimensional Minkowski spacetime:
the statistics in general is a representation of the braid
group, related (via the Spin-Statistics theorem) to a much wider range
of helicities than the Fermi-Bose alternative in $D=4$.  

In contrast to $D>2$ dimensions, the conformal group in two spacetime
dimensions (2D) turns out to be infinite-dimensional. In fact, the
Dirac manifold in two dimensions can be identified with the product of
two ``chiral'' circles $S^1\times
S^1$ on which the conformal group $\Cf_2$ acts as a product of two
infinite-dimensional Lie groups $\Diff_+(S^1)$ (the
orientation-preserving diffeomorphisms).  

Namely, one can parameterize the solutions to $\xi\cdot\xi=0$ for
$\xi\in\RR^{2,D}$ as
$\xi=\lambda\cdot(\sin\alpha,\sin\beta,\cos\beta,\cos\alpha)$, so that 
$[\xi]\leftrightarrow
(z^+=e^{i(\alpha+\beta)},z^-=e^{i(\alpha-\beta)})$ is a bijection
between the quotient manifold and the product $S^1\times S^1$ of two ``chiral
circles''. Moreover, the chiral Minkowski coordinates are
\be
x^\pm\equiv x^0\pm x^1=\frac{\xi^0\pm\xi^1}{\xi^2+\xi^3}=\frac{\sin\alpha\pm\sin\beta}{\cos\beta+\cos\alpha},
\ee
such that
\be \label{cayley}
z^\pm=\frac{1+ix^\pm}{1-ix^\pm}.
\ee
This is the Cayley transform $\RR\to S^1$ mapping the light ray into
the circle (excepting the point $z=-1$), whose inverse is the
stereographic projection $S^1\setminus\{-1\}\to\RR$. 
Because the Minkowski metric factorizes as $ds^2 = dx^+dx^-$, the
independent diffeomorphisms of the chiral circles preserve the
metric up to a factor. 

As it turns out, nontrivial unitary positive-energy representations of
the group $\Diff_+(S^1)$ are necessarily projective representations,
and a state invariant under $U_g$ for all diffeomorphisms $g$ does not
exist. As a consequence, the vacuum vector of a conformal QFT in two 
dimensions is invariant only under the product of the two M\"obius groups
$\SL(2,\RR)/\ZZ_2\sim \SU(1,1)/\ZZ_2\subset \Diff_+(S^1)$. 
This unbroken subgroup of $\Cf_2= \Diff_+(S^1)\times\Diff_+(S^1)$ coincides with the group
$\SO(2,2)_0/\ZZ_2$ (which is the conformal group $\Cf_D$ in $D>2$
dimensions if one puts $D=2$). 

An important feature of 2D conformal QFT (also related to the
factorization of the metric) is the presence of chiral observables, 
which depend only on $x^+$, or on $x^-$. These fields are therefore
defined on the circle $S^1$, into which $\RR$ is
embedded via the Cayley map. The M\"obius group acts by
fractional linear transformations $\PSL(2,\RR)$ 
on $\RR\cup\{\infty\}$ or by fractional linear transformations
$\PSU(1,1)$ on $S^1$: 
$$x\mapsto\frac{ax + b}{cx+d}\,,\quad \bpm a & b \\ c & d
\epm\in \SL(2,\RR), \qquad z\mapsto\frac{\alpha z + \beta}{\ol \beta z+\ol
  \alpha}\,, \quad \bpm\alpha & \beta \\ \ol\beta & \ol\alpha \epm\in
\SU(1,1).$$  

The presence of chiral fields is strongly connected with conservation
laws. Because a symmetric tensor field that transforms irreducibly
under the conformal group is traceless, it has only two independent
tensor components $T_{+\dots+}$ and $T_{-\dots-}$ (where
$\pm$ stands for Lorentz indices $0\pm 1$). If a symmetric
tensor current is conserved, the continuity equation implies that
these components are chiral fields, i.e., $\partial_-T_{+\dots+}=0$
and $\partial_+T_{-\dots-}=0$.  
(This simple argument fails for rank $1$, because the current
conservation law $\partial_\mu J^\mu=0$ gives only one equation. But
one notes by inspection that the unique conformally covariant
two-point function of a conserved current is also dually conserved,
hence by the Reeh-Schlieder theorem also the dual current is
conserved: $\partial_\mu \,\eps^{\mu\nu} J_\nu=0$. Then again, $J_+$
and $J_-$ are chiral fields.)  
Notice also that the conservation of the conformally invariant
two-point function fixes the scaling dimension $d$ of a conserved
tensor current, to coincide with the tensor rank $r$.

Conversely, every pair of chiral fields of equal dimension constitutes
a conserved traceless symmetric tensor current of rank $r=d$ as a
two-dimensional tensor field. Thus, chiral fields naturally occur in
2D conformal QFT models whenever this theory has local conservation
laws or, by Noether's 
theorem, continuous symmetries, and they actually represent the local
generators of these symmetries. 

The most important conserved tensor current is the stress-energy
tensor (SET; dimension $=$ rank $=2$), which is (by definition) the local
generator of the conformal symmetry $\Diff_+(S^1)\times \Diff_+(S^1)$
itself. The L\"uscher-Mack theorem \cite{LM} fixes the self-commutator
of its chiral components up to the ``central charge'' $c$, which can be
regarded as the coefficient of the unique central extension of the Lie
algebra of the diffeomorphism group of the circle. This quantity is a
distinctive invariant of the chiral CFT at hand (where a priori,
unless the 2D theory is parity symmetric, the two chiral central
charges need not to coincide).  

By local commutativity in the two-dimensional Minkowski spacetime, $+$
and $-$ chiral fields (``left- and 
right-movers'') commute with each other, and chiral fields of the same
chirality commute at non-coinciding points, so that their commutators
are linear combinations of other fields multiplied with (derivatives of)
$\delta$-distributions in the chiral variables $x^\pm$. 

Any algebraic relations among chiral fields, including their local
commutators, are strongly restricted by conformal covariance. 
Early attempts at classification of chiral CFTs tried to find consistent
commutation and operator product relations (``$W$-algebras''); but in this approach it is
in general not known how to assess the existence of representations on a
Hilbert space. The most prominent cases where this is possible, are
the quantizations of the central charge $c$ below 1 \cite{FQS}, and of 
the ``level'' $k$ of chiral current algebras \cite{K}, which arise precisely due
to the unitarity of the vacuum representation. Both these
classification results are made possible by the fact that local commutation
relations of the stress-energy tensor and of current algebras can be
regarded as central extensions of infinite-dimensional Lie algebras
(the Virasoro algebra, and affine Kac-Moody algebras, respectively),
which are obtained by extending the fields as operator-valued
distributions on $S^1$ and Fourier-decomposing the latter. Then,
methods of highest-weight representations of Lie algebras can be
applied to obtain the mentioned results. 

The most basic examples of chiral fields are free fields related to
the 2D massless Dirac and Klein-Gordon fields, which can be
constructed in a standard way in terms of creation and annihiliation
operators on a Fock space: 

$\bullet$ The massless 2D Dirac Fermi field decouples into its two
chiral components, where ``chiral'' stands for the projections onto
the eigenvalues of the Dirac matrix $\gamma^5=\gamma^0\gamma^1$ which
fix the sign of the helicity relative to the momentum of the
massless particles states; by virtue of the massless Dirac equation,
the chiral components depend only on $x^+$ or $x^-$, respectively. 
(While this original meaning of the term ``chiral'' refers to the helicity,
we shall below understand it in the sense of ``dependence on 
$x^+$ or $x^-$ only''.)  

$\bullet$ The canonical massless Klein-Gordon field is ill-defined as an
operator-valued distribution on all test functions in $\mathcal{S}(\RR^2)$. It
is, however, well-defined on test functions that are derivatives of test
functions -- which is tantamount to considering the ``gradient of the
Klein-Gordon field'' as an operator-valued distribution on all test
functions; this gradient field is of course a conserved current. 
Its chiral components are chiral fields of scaling dimension $d=r=1$. 

The stress-energy tensors of these free fields are not free fields
themselves, but can be written as Wick products of the free fields. 
The SET of a real chiral Fermi field $\psi$ has $c=\frac12$, that of
the chiral current $j$ has $c=1$. A remarkable feature arises here,
known as ``fermionization'': The chiral current, which is a free
Bose field, is at the same time a neutral Wick product $\wick{\psi^*\psi}$
of a complex free Fermi field (= two real free Fermi fields), acting
as the local generator of its $U(1)$ gauge symmetry $\psi(x)\mapsto
e^{i\alpha(x)}\psi(x)$. Likewise, the SET of the Bose current
$T=\pi\,\wick{j^2}$ coincides with the SET of the complex Fermi 
field, $T=\frac i2\wick{\psi^*\partial\psi-\partial\psi^*\psi}$, 
so that the same field has two representations in terms of free Bose
or free Fermi fields. 

(For the more remarkable converse, called ``bosonization'',
i.e., the representation of fermionic fields in terms of bosonic
fields, see \sref{s:boso}.) 

Writing ``the same field'' in the previous exposition means, that all
vacuum correlation functions coincide in both representations. But the
vacuum Hilbert space of the SET is just a subspace of the Fock space
of the current, which is in turn a subspace of the Fock space of the
complex chiral Fermi field. Thus, the latter Fock space carries
reducible representations of the current and of the SET.

This simple example leads us to the issue of {\em representations}.

\subsection{Representations}

The specific algebraic relations defining a QFT in general admit
many inequivalent Hilbert space representations. Covariant
representations $\pi$ of the field algebra come with a (projective) unitary representation
$U_\pi$ of $\Cf_D$ whose adjoint action on the covariant fields implements the
transformation law: 
\be
U_\pi(g)\pi(\phi(x)) U_\pi(g)^* = \pi(\alpha_g(\phi(x))) =
D(g,x)\inv\pi(\phi(gx)).
\ee

A positive-energy representation is a covariant representation of the field algebra in which
the generator of the unitary one-parameter group of time translations
(the Hamiltonian) has positive spectrum: $P_0\geq 0$. This implies
that the commuting generators $P_\mu$ of the subgroup of translations
have joint spectrum in the closed forward lightcone ($P\cdot P\geq 0,
P_0\geq 0$). A vacuum representation has in addition a unique ground state
of zero energy, $P_0\Omega =0$.

In conformal QFT, positivity of the Hamiltonian $P_0$ is equivalent to
positivity of the generator $K_0=U(I)P_0U(I)^*$ of the timelike
special conformal transformations, and of the ``conformal
Hamiltonian'' $H_{\rm conf}=\frac12(P_0+K_0)$. In 2D, positivity
of the generators of chiral M\"obius transformations  
$P^\pm=P_0\pm P_1$, $K^\pm=K_0\mp K_1$ and $L_0^\pm = \frac12
(P^\pm+K^\pm)$ follows. $L^\pm_0$ are the generators of the ``rotations''
of $S^1$. 

In the free field example above, the Fock space of the complex Fermi
field splits into an infinite direct sum of charged positive-energy
representations of the current (the neutral representation being the
vacuum representation); in turn, as a representation of the SET, the
vacuum representation of the current 
decomposes into an infinite direct sum of representations of the $c=1$ 
Virasoro algebra. The precise decompositions can be read off the
``chiral characters'': these are the power series $\chi(t)=\Tr
\exp(-\beta L_0)$ or $\chi(t,q)=\Tr
\exp(-\beta L_0-\mu Q)$ in the variables $t=e^{-\beta}$ and
$q=e^{-\mu}$, which yield the multiplicities of the (integer or
half-integer) eigenvalues of the conformal Hamiltonian $L_0$ and
charge operator $Q$ in the respective representation spaces.

The classification of the positive-energy representations of the
Virasoro algebra and of affine Kac-Moody algebras \cite{FQS,K} is a breakthrough
for QFT without precedent, and without analog in $D=4$: not only are
the algebras generated by the stress-enery tensor, resp.\ by currents
of dimension $1$, universally fixed by conformal invariance (up to the
central charge in the first case, and up to the structure constants of
a Lie algebra and the level in the latter case), but also all their
positive-energy representations are explicitly known without a
perturbative construction. These models stand at the beginning of many
remarkable findings in the representation theory of more general
chiral CFTs, including modular tensor categories and the Verlinde formula. 

Many issues in QFT are of a basically representation theoretic nature;
e.g., the vacuum representation of an extension of a given QFT is in
general a reducible representation of the latter, and certain data
pertaining to this representation can be used to classify
extensions. (An ``extension'' is, broadly speaking, a QFT containing a given
QFT with the same stress-energy tensor as generator of the covariance;
see \sref{s:Qsys} for more details.) 
Another issue are QFTs with boundaries, whose boundary conditions can
be understood as different representations of a suitable quotient
algebra \cite{BPB}, see \sref{s:phase}. These topics can be nicely
addressed in two-dimensional CFT and give rise to several nontrivial
classifications, because the structure of superselection sectors of
chiral CFT is well understood.   

In particular, a two-dimensional CFT is an extension of the tensor
product of a pair of chiral CFTs; but also many chiral CFTs can be
constructed as extensions or subtheories of other chiral CFTs (see
\sref{s:Qsys}).  

\subsection{Different axiomatizations}

Representation theoretic issues are best captured in the algebraic
formulation of QFT (AQFT), emphasizing QFT as a given algebraic structure
that admits many inequivalent Hilbert space representations. The
Haag-Kastler axioms of AQFT therefore do not presume the existence of
a vacuum vector -- its presence is rather a feature of the
representation considered. 

Apart from this, the main difference to the Wightman axiomatics is
that the same physical principles are reformulated in terms of local
observables rather than quantum fields, thus offering a somewhat
broader generality. We shall briefly review this approach in
\sref{s:aqft}, and be more detailed for chiral conformal QFT in
\sref{s:acft}, where we also present its most important general
results.   

It follows from these general axioms that the structure of
superselection sectors ($=$ positive-energy representations) of
completely rational (see \sref{s:acft}) models of chiral CFT is captured
by a modular C* tensor category. The latter is therefore the basic
tool to describe the algebraic structure of two-dimensional CFT models
(as extensions of chiral ones). There is a wealth of abstract
mathematical results about braided and modular tensor categories. We
shall point out in \sref{s:Qsys} and \sref{s:phase} that many of these
abstract results have natural algebraic and representation-theoretic
counterparts in the setting of chiral and two-dimensional CFT.  

This nearly perfect match is the reason why certain popular
axiomations of CFT actually ``start from the other end'', taking a
modular tensor category as the initial axiomatic data. This point of
view is justified by the above line of argument, but it tends to
create the impression that conformal QFT, rather than being a special case
of relativistic QFT, were a ``world of its own'', with all its peculiar
features -- notably the presence of a braided or modular tensor
category -- being not only {\em admitted} by the general axiomatic 
frameworks, but being actually {\em consequences} of the usual 
axioms, specialized to 2D and augmented by conformal symmetry. 

Especially, modularity of the representation category should not be
regarded as an axiom reflecting some fundamental physical principle, 
since important models, like the $u(1)$ current algebra, 
do not share this property. As the characterization in \cite{KLM}
shows, it follows from complete rationality which is rather a
regularity property than a fundamental feature.

Yet different axiomatisations, e.g., Euclidean CFT or vertex operator
algebras (VOA), also capture essential features of rational CFT, without
requiring all the features of relativistic QFT. Notably, the Hilbert
space axiom is not essential in some approaches, which therefore
admit even more classes of models. Euclidean CFT has a direct physical
interpretation independent of its possible correspondence to a 
real-time relativistic CFT, as critical limits of classical lattice
systems in two space dimensions (even with experimental
verification). The physical interpretation of ``the most general VOA''
is not known, but with additional assumptions a close tie with
relativistic conformal QFT can be established \cite{CKLW}.

\section{Algebraic QFT}
\setcounter{equation}{0}
\label{s:aqft}
The algebraic approach to quantum field theory is formulated by the
Haag-Kastler axioms \cite{H}. The localization of local observables is
captured not by the use of quantum fields, but rather by specifying
``local subalgebras'' of a suitable global C* algebra. To each open
spacetime region is associated a local algebra, whose self-adjoint
elements are supposed to represent the physical observables that can
be measured (or operations that can be performed) in that region:
$$O\mapsto A(O).$$ 
Unitary exponentials or spectral projections of self-adjoint smeared
field operators would generate local algebras, but this kind of
construction is not necessarily assumed in the AQFT approach. It is a
nontrivial challenge to find criteria to decide whether local algebras
come from Wightman fields, and to extract the latter from the former
\cite{BH}.  

The principles of covariance and causality are easily formulated: The
group of spacetime symmetries acts by automorphisms $\alpha_g$ of the
global algebra properly transforming its local subalgebras into each
other, and the local algebras associated with two spacelike separated regions
are mutually commuting subalgebras.

Appropriate spacetime regions in Minkowski spacetime are doublecones $O$
(intersections of past and future lightcones). With respect to the
order relation among doublecones by inclusion, the assignment
$O\mapsto A(O)$ is a ``net'' of local algebras. (Since the set of
doublecones in the Dirac manifold is only partially ordered, the local
algebras constitute only a pre-cosheaf of algebras.) 


By postulating the local algebras to be C* algebras, and the ``global''
algebra to be the C*-inductive limit of the local algebras on
Minkowski spacetime (usually called the quasilocal algebra), it is
ensured that there exist Hilbert space representations. Among all
Hilbert space representations, one should select those which describe
states of physical interest. This accounts for the fact, that quantum
field theory -- unlike quantum mechanics -- admits in general many
inequivalent representations, among which positive-energy
representations (distinguished by the implementation of time
translations by a unitary one-parameter group with positive generator)
and thermal equilibrium representations (distinguished by the presence
of a KMS state ensuring the appropriate thermodynamic stability and
passivity properties) are the most important ones \cite{BH}. Our focus
in the sequel will be only on positive-energy representations. 

We shall describe various ways of specifying local algebras
in \sref{s:chimod}.  

One may object that a ``net of local algebras'' is a structure too
abstract for specifying a particular model -- in particular, it does not
involve an explicit specification of a Lagrangian. But recall that
a model of Quantum Mechanics is fixed by specifying the set of
observables on the Hilbert space (typically all selfadjoint operators)
and the Hamiltonian, i.e., the time evolution automorphisms
$\alpha_t=\Ad_{U(t)}$, $U(t)=e^{iHt}$. The same is true in QFT: one
has to specify the algebras of observables and the relativistic covariance.  


It is a crucial fact in this respect that scattering theory can be
carried out (at least in positive-energy representations with a mass
gap), by constructing multi-particle states and the scattering matrix,
using only the net of local algebras and its covariance.  
Thus, the local algebras ``contain'' all the information that
is needed to provide the 
interpretation of a model in terms of
particles and their interactions. This is 
in accord with the fact that, in collider experiments, one usually does
not measure a particular field strength but rather deposits of
``something'' in the detector arrays, and the physical interpretation
of this ``something'' as a particle of either kind is imposed by the
correlations of signals in different detector cells (naturally
interpreted as ``particle tracks'' ). In Haag-Ruelle scattering
theory, the asymptotic dynamics of any local observable applied to the
vacuum state (as long as it is not orthogonal to the desired particle
state) is sufficient to identify the asymptotic particle states \cite{H}. 

The specific dynamics of course enters through the specification of
the time evolution automorphisms as part of the covariance. Thus, the
Lagrangian (if there is any) is hidden as an ``inverse scattering
problem'' in the scattering theory of the net of local algebras. 

There is a marked difference to the standard approach (quantization of a
classical Lagrangean theory): it is not a priori required that the
generators of spacetime symmetries (in particular the Hamiltonian) are
integrals over densities (components of the stress-energy tensor) which
are some local ``functions'' of the field observables. In the
classical theory, such relations would imply, through the
canonical Poisson brackets, the correct infinitesimal
transformation laws on the fields. In quantum field theory, on the other
hand, canonical commutation relations cannot hold (in general) in a
strict sense. Retaining only the ``correct infinitesimal
transformation laws'' is therefore an appropriate substitute for the
quantization of a classical Lagrangean formulation. One should keep in
mind that classical physics is only a limit of the
``true'' quantum physics, and there is no reason to believe in a
fundamental 1:1 correspondence between the two realms. 


The ``lack of fields'' in the AQFT framework is in fact another strength: 
while different sets of quantum fields (relatively local w.r.t.\
each other) may generate the same local algebras, and hence have the
same scattering states and the same scattering matrix (thus they
belong to the same ``Borchers class''), the local algebras they
generate are the same, and should be regarded as the invariant content
of the theory. The actual choice of fields may rather be regarded as
an auxiliary device to describe the algebras (analogous to the choice
of coordinates for a manifold), which may be very convenient but is
not intrinsic to the physical interpretation. 


In the application of the AQFT framework to conformal QFT in two
dimensions, nets of local algebras are not just {\em assumed} to be
given (as often in axiomatic  
approaches), but can be explicitly constructed by a large variety of
methods. Some of these models actually do use fields (e.g., free Fermi
fields which are bounded operators after smearing, or currents whose
unitary Weyl operators are taken as the generators of the local
algebras), but manipulations on such elementary constructions give
rise to new nets that cannot always be easily described in the
language of fields. We shall contrast the usual ``field-theoretical''
construction methods with the algebraic methods in \sref{s:chimod}.


Finally, it is a mathematical benefit that one may work with bounded
operators, with norm given by the C* structure, and has not to worry
about domains of definition. In this setup, it is easy to say what a
``representation'' is, and the issue of superselection sectors as
unitary equivalence classes of positive-energy representations can be
addressed. Algebraic quantum field theory is therefore the ideal setup
to approach representation-theoretic issues. 

Most of the seminal breakthrough achievements in conformal
quantum field theory in two spacetime dimensions are of a
representation theoretic nature (classification of the central charge
and of conformal dimensions at $c<1$ \cite{FQS}, fusion rules
\cite{BPZ}, coset models and branching rules \cite{GKO}, \dots). 
Indeed, AQFT provides a unifying framework for these insights, that
has proven one of the places where the AQFT formulation is most powerful,
by the general theory of superselection sectors initiated by 
Doplicher, Haag and Roberts \cite{DHR}, see \sref{s:DHR}.

\section{Algebraic CFT on the circle}
\setcounter{equation}{0}
\label{s:acft}

\subsection{Axioms}
\label{s:axioms}

We give here the AQFT axioms for a M\"obius covariant chiral QFT
(chiral CFT) directly in its vacuum representation, as in \cite{BGL93}.

As emphasized earlier, in order to be prepared for the study of more
general representations, one should rather axiomatize a pre-cosheaf of
abstract local algebras on which the M\"obius group acts by
automorphisms; but one may as well 
read off this pre-cosheaf from its vacuum representation, by regarding
the latter as the defining representation.

A chiral CFT is thus given by a family of local algebras $A(I)$ on a
Hilbert space $H$ where $I$ runs over the proper open intervals of the
circle $S^1$ into which the real axis $\RR$ is embedded via the Cayley
transform \eref{cayley}. One may take $A(I)$ to be von Neumann algebras. The
fundamental axioms are: 

(1) {\bf Isotony.} If $I_1\subset I_2$, then 
\be
A(I_1) \subset A(I_2).
\ee

(2) {\bf Locality.} If $I_1 \cap I_2 = \emptyset$, then 
\be
[A(I_1),A(I_2)] = 0.
\ee

(3) {\bf M\"obius covariance.} There is a unitary representation $U$
of the M\"obius group $\PSU(1,1)=\PSL(2,\RR)$ on $H$ such that 
for any interval $I$ and $g \in \PSU(1,1)$,  
\be
\alpha_g\big(A(I)\big):= U_g \pi_0(A(I)\big)U_g^* = A(gI).
\ee

(4) {\bf Positive energy.} 
The generator $P$ of the one-parameter
subgroup of translations in the representation $U$ has positive spectrum.

(5) {\bf Vacuum.} There is unique (up to a phase) unit vector
$\Omega\in H$ which is invariant under the action of $U$, and
cyclic for $\bigvee_IA(I)$.

As mentioned above, positivity of the chiral Hamiltonian $P$ is
equivalent to positivity of the generator $L_0$ of the one-parameter
subgroup of rotations (the conformal Hamiltonian). By (4) and (5), the
spectrum of $L_0$ in the vacuum representation is a subset of $\NN_0$.

The axioms (1)--(5) imply the following properties \cite{BGL93}:

(6) {\bf Reeh-Schlieder property.} The vector $\Omega$ is cyclic and
separating for each local algebra $A(I)$. 

(7) {\bf Additivity.} If $I = \bigcup_i I_i$ , then $A(I)=\bigvee_i
A(I_i)$. 

(8) {\bf Haag duality on $S^1$.} For any proper interval $I$ one has 
\be
A(I) = A(I')',
\ee 
where $I'$ is the interior of the complement of $I$ in $S^1$.

(9) {\bf Bisognano-Wichmann property.} The Tomita modular group \cite{MT} 
$\Delta^{it}$ of $A(\RR_+)$ with respect to the vector $\Omega$
coincides with $U(\delta_{-2\pi t})$, where $\delta_t\in \PSL(2,\RR)$ is the
one-parameter group of dilations. 

(Here and below we are freely using the Cayley identification of $\RR$
as a subset of $S^1$, and $\PSU(1,1)=\PSL(2,\RR)$. By virtue of
M\"obius covariance, the analog of (9) is true for every local algebra
$A(I)$ and the associated subgroup of ``dilations'' of $I$.)  

An interesting consequence of (9) is that the modular groups
of the local algebras of any three intervals generate the
representation $U$ of the M\"obius group $\PSU(1,1)$, in particular,
spacetime symmetries are of modular origin. Conversely, it was
shown in \cite{GLW} that if $M_i$ are three commuting von Neumann algebras
with a joint cyclic and separating vector $\Omega$, and $M_i\subset
M_{i+1 \mod 3}'$ are halfsided modular inclusions, then the three
modular groups generate a positive-energy representation $U$ of
$\PSU(1,1)$. From this, one can construct a conformal net in its
vacuum representation satisfying the axioms (1)--(5) (and strong
additivity (12) below) by identifying $M_i$ with the local algebras of
three intervals arising by subdividing the circle by 
removing three points, and their modular groups with the corresponding
dilation subgroups of $\PSL(2,\RR)$, and using the action of the resulting
representation $U$ to consistently define $A(I)$ for general $I$.  

(This seems to be an interesting way to construct new chiral CFT
models, but ideas to provide triples of commuting von Neumann algebras
with the stated properties are scarce if one does not {\em start}
from a CFT.)

Beyond the basic axioms (1)--(5), one may require further properties 
that are satisfied for many models. 

The presence of a stress-energy tensor is axiomatized as a
stronger version of (3):

(10) {\bf Diffeomorphism covariance.} The representation $U$ of
$\PSU(1,1)$ extends to a projective unitary representation of
$\Diff_+(S^1)$ 
such that for any interval $I$, one has
\be
\Ad_{U(g)}A(I) = A(gI), \quad\hbox{for}\quad g \in \Diff_+(S^1),
\ee
and 
\be
\Ad_{U(g)} a = a, \quad\hbox{if}\quad a\in A(I),\,\, \supp g \subset I'.
\ee
By Haag duality (8), the latter property implies that $U(g)\in A(I)$ if
$\supp g\subset I$, and the subnet generated by such $U(g)$ is called a
Virasoro net. 

The next two properties express ``maximal decoupling'' of local
observables in intervals at a finite distance, and ``maximal
interaction'' of local observables in touching intervals \cite{LX}:  

(11) {\bf Split property.} If $I_1$ and $I_2$ are two intervals with
disjoint closure, then the map $a_1\otimes a_2\mapsto a_1a_2$ is an
isomorphism of 
von Neumann algebras
\be
A(I_1)\otimes A(I_2)\simeq A(I_1)\vee A(I_2).
\ee

An equivalent assertion is that states can be independently prepared on
the subalgebras $A(I_1)$ and on $A(I_2)$, such that a joint state
restricting to the given states on $A(I_i)$ always exists. This
feature depends on the energy level density, and is known to be true
\cite{BW} if $e^{-\beta L_0}$ is a trace-class operator (in the vacuum
representation) for every $\beta>0$. (Such traces, regarded as power
series in $t=e^{-\beta}$ whose coefficients give the multiplicities of 
the spectrum of $L_0$, are usually referred to as ``characters'' and
are a very useful tool for the decomposition of reducible representations.) 
 
(12) {\bf Strong additivity.} Whenever two intervals $I_1$ and $I_2$
are obtained by removing an interior point from a proper interval $I$, then 
\be 
A(I) = A(I_1) \vee A(I_2).
\ee

Thinking in terms of quantum fields, (12) may be regarded as a
regularity property to the effect that the smearing can be
approximated by test functions that vanish at a given point. 

In view of (8), strong additivity (12) is equivalent to Haag duality 
on $\RR$, namely for any proper interval $I\subset \RR$ one has 
\be
A(I) = A(I^c)',
\ee
where $I^c$ is the interior of the complement of $I$ in $\RR$.

For the algebra of two intervals $I,J\subset S^1$ with disjoint closure, 
Haag duality will generally fail. The
index of the inclusion (in the vacuum representation) 
\be
A(I\cup J)\equiv A(I)\vee A(J)\subset A\big((I\cup J)'\big)'
\ee
is called the $\mu$-index (which my be infinite). 

A chiral CFT is called ``completely rational'' \cite{KLM,L03,LX} if it
is split (11), strongly additive (12), and has finite $\mu$-index. A
chiral CFT is called ``rational'' if it possesses only finitely many
irreducible superselection sectors, see \sref{s:DHR}. It was shown in
\cite{LX} that a split chiral CFT is completely rational if and only
if it is rational; in particular, rationality together with split implies
strong additivity. Moreover, in a completely rational chiral CFT the
braiding is completely non-degenerate, turning the C* category of
superselection sectors (see \sref{s:DHR}) into a modular category \cite{KLM}.  

It should be stressed, however, that complete rationality is by no
means an obvious feature. One of the most elementary models, the
chiral $u(1)$ current algebra, satisfies the split property and strong
additivity, but possesses a continuum of charged sectors, and thereby
fails to be rational, hence completely rational.

\subsection{Superselection sectors}
\label{s:DHR}

A general theory of superselection sectors was originally developped
by Doplicher, Haag and Roberts \cite{DHR} to describe sectors in
massive QFT in four spacetime dimensions. With minor modifications in
the setup, but big differences in the outcome (see below), it has
proven to be applicable in two-dimensional and chiral CFT \cite{FRS}.

The crucial assumption on the representations $\pi$ that can be
treated by the DHR theory of superselection sectors is that, upon
restriction to the causal complement of any doublecone region, $\pi$
is indistinguishable from the vacuum representation up to unitary equivalence:
\be
\label{crit}
\pi\big\vert_{A(O')}\simeq \pi_0\big\vert_{A(O')}.
\ee
The heuristics behind this criterion is the argument that a charge
distinguishing two representations can always sit inside the
doublecone $O$, which is inaccessible to measurements in its causal
complement $O'$. 

Even for massive theories, this heuristic idea may fail for charges 
that can only be localized in ``topological strings'' (narrow cones
extending to spacelike infinity) \cite{BF}, requiring some mild
adaptation of the theory. More dramatically, the criterium \eref{crit}
excludes theories with long-range forces, notably QED, because an
electric charge can be detected by measurements in the causal
complement due to Gau\ss' law. An adaptation of the theory to this
case has recently been formulated by Buchholz and Roberts \cite{BR}.

In contrast, the chiral analogue of \eref{crit} is automatically
satisfied in chiral CFT \cite{BMT}. 

Assuming the validity of the criterium \eref{crit}, positive-energy
representations $\pi$ can be described in terms of DHR endomorphisms
$\rho$ of the quasi-local algebra \cite{DHR}. (In the case of CFT on
$S^1$, where the set of intervals is only partially ordered, one may
restrict the CFT to a net on $\RR\subset S^1$. Otherwise, the definition of DHR
endomorphisms is slightly more involved, invoking the pre-cosheaf
structure.) Up to unitary equivalence, one has   
\be
\pi=\pi_0\circ\rho,
\ee
where $\pi_0$ is the (defining) vacuum representation.  

The DHR endomorphisms are ``localized'' in some region, meaning that
their action is trivial on the subalgebras of local observables at
spacelike distance of that region, and ``transportable'', meaning that
that region can be chosen arbitrarily. The first property is
consistent with the idea that $\rho$ arises by conjugation
with some localized charged field operator. The second is
(a consequence of) covariance, and is consistent with the idea that
the total charge does not depend on the localization of the charged
field operator. But it should be stressed that these properties are derived
{\em without assuming} the existence of such a field operator, and are
entirely intrinsic in terms of the observables. 

Equivalence, direct sums and subrepresentations of positive-energy
{\em representations} can be formulated directly for the
corresponding DHR {\em endomorphisms}, in terms of intertwiners, which
are local observables satisfying the intertwiner relation 
$t\rho_1(a) = \rho_2(a)t$. The crucial insight of \cite{DHR} is that
locality, Haag duality and the localization of DHR endomorphisms imply
algebraic properties of the intertwiners, which turn the
representation theory into a C* tensor category (the ``DHR category'').   

The composition of DHR endomorphisms describes a product of
representations (``fusion'') with the vacuum representation (the
identical endomorphism) as the ``neutral'' element. The fusion product
is commutative up to 
unitary equivalence, implemented by distinguished unitary intertwiners 
(``statistics operators'') \cite{DHR,FRS}. The statistics operators
turn the DHR category into a braided tensor category. In particular,
each irreducible sector can be assigned a ``statistics phase''
$\kappa_\rho$ and a ``statistical dimension'' $\dim(\rho)$. 

In four dimensions, the braiding is in fact a permutation
symmetry. As a consequence, $\kappa_\rho=\pm 1$ and $\dim(\rho)\in
\NN$, and these quantum numbers are related to the statistics (and
hence the spin) of particles in the associated charge sector 
\cite[II]{DHR}. In chiral CFT, the conformal spin-statistics theorem
\cite{GL96} relates the statistics phase with the conformal spin,
namely the value of the unitary representative $U(2\pi)$ of the 
full rotation of $S^1$, and hence the spectrum of the generator $L_0$ (mod $\ZZ$). 

In a large class of chiral CFT models (the completely rational ones,
see \sref{s:axioms}), the braiding is in fact maximally
non-degenerate, so that the DHR category is even a modular
C* tensor category \cite{KLM}, and the sum $\sum_\rho \dim(\rho)^2$
over all irreducible sectors equals the $\mu$-index, see \sref{s:axioms}.  

The latter is a very interesting result, quantitatively relating the
existence of nontrivial superselection sectors to a quantity that can
be ``measured'' in the vacuum representation. 

The large variety of available models opens the door to model studies
of general concepts, exploring the range of possibilities admitted by
the general principles of local quantum field theory. Although these
methods are presently limited to two dimensional conformal QFT,
viewing them in the context of general QFT may be instructive for 
the construction of more realistic models in four-dimensional
spacetime. We shall discuss some of these issues in \sref{s:algebraic}.

\section{Chiral model constructions}
\setcounter{equation}{0}
\label{s:chimod}
\subsection{Free fields}
\label{s:free}

The most elementary constructions of chiral models are, as always,
free fields. As the conformal scaling dimension $h$ specifies the
M\"obius invariant two-point function 
$\propto \big(-i/(x-y-i\eps)\big)^{-2h}$, and the two-point function
completely determines a free field, the only ``choice'' is a scaling
dimension $h$ which must be positive in order that the two-point
function is a positive-definite scalar product. The two-point function
is local iff $h\in \NN$, and it is anti-local iff $h\in \frac12+\NN_0$. 

The case $h=\frac12$ is identical with the chiral component of a
massless Majorana resp.\ Dirac field, which are real resp.\ complex
chiral Fermi fields. The case $h=1$ is identical with the ``chiral 
derivative of the massless Klein-Gordon field'', called the free
current, see \sref{s:2D}.

In the AQFT framework, one would rather define the chiral free
Fermi field as a CAR algebra over $L^2(S^1)$ with the vacuum state
$\omega\big(\psi(f)\psi(g)\big)=(\ol f,\Pi_+g)$ specified by the projection
$\Pi_+$ onto the positive-frequency part; and the 
chiral free current can be defined by the CCR algebra over $C(S^1)$
with symplectic form $\sim i\int (f'g-fg')$ with the vacuum state
given by a Gaussian $\omega(W(f))=e^{-\frac14\norm{f}_C^2}$ with a
suitable inner product of the complexified symplectic space to
guarantee positive energy. Local subalgebras are specified by
specifying the support of the functions $f$.

\subsection{Wick products and Fermionization}
\label{s:fermi}

Quadratic Wick products of free fields are well-defined by standard
quantum field theory methods.

In particular, the stress-energy tensor of the free Fermi field is
$T=\frac i2\wick{\psi\partial\psi}$, and the stress-energy tensor of a
free current is $T=\pi\,\wick{j^2}$, giving the central charge
$c=\frac12$ and $c=1$, respectively.

``Fermionization'' is the remarkable feature that the bosonic free
current can be represented as the neutral Wick product
$\wick{\psi^*\psi}$ of a complex free Fermi field. The bosonic current is
therefore defined on the fermionic Fock space, which as a
representation splits into an infinite direct sum of charged
representations with integer charge. The bosonic stress-energy tensor
$\pi\,\wick{j^2}$ coincides with the fermionic stress-energy
tensor $\frac i2\wick{\psi^*\partial\psi-\partial\psi^*\psi}$. 

By way of generalization, one obtains ``nonabelian currents''
associated with a (semi-simple) Lie algebra $\Lg$ by the ``quark model
construction'': choosing an $n$-dimensional real or complex matrix
representation $\tau^a=\pi(X^a)$ of the generators $X^a$ of $g$, one
defines the currents as quadratic Wick products of $n$ real or complex
free Fermi fields: $j^a\sim\tau^a_{ij}\wick{\psi_i\psi_j}$
resp.\ $j^a\sim\tau^a_{ij}\wick{\psi_i^*\psi_j}$. 

Their commutation relations can be viewed as central extensions of the
Lie algebra $L\Lg$ of the loop group $LG$, i.e., the $\Lg$-valued
functions on $S^1$; the central term is universal up to a factor,
called the level $k$, and the level in the given construction is a
function of the Lie algebra $\Lg$ and its representation $\pi$.

The resulting currents $j^a(f_a)$ act as infinitesimal generators of 
gauge transformations $\ul\psi(x)\mapsto e^{-if_a(x)X^a}\ul\psi(x)$. 

Given a nonabelian current algebra, its stress-energy tensor is given
by the Sugawara construction, $T_S\sim h_{ab}\wick{j^aj^b}$ where
$h_{ab}$ is the invariant Killing metric, and the normalization factor
is determined by the Lie algebra $\Lg$. It should be noted that if the
currents are obtained by the ``quark model construction'' on a
fermionic Fock space, the Sugawara stress-energy tensor for the
currents will in
general not coincide with the fermionic stress-energy tensor (see
\sref{s:coset}).  

This construction has an analogue in the AQFT framework \cite{PS,W}. Let
$G$ be a compact Lie group with a unitary representation $\pi$ on
$\CC^N$. One notes that the gauge transformations
$\alpha_g:\ul\psi(x)\to \pi(g(x)\inv)\ul\psi(x)$ for $G$-valued
functions $g$ on $S^1$, i.e., $g\in LG$, are automorphisms of the CAR
algebra of $N$ Fermi fields. The criterium for
implementability by unitary operators $W(g)$ in the GNS-Hilbert space
of the vacuum state (i.e., the Fock space) can be verified to be
fulfilled. It follows that the unitaries $W(g)$ define a projective
representation of the loop group $LG$. The cohomology class of the
cocycle of this representation can be identified with the level, such
that, up to a coboundary, $W(g)=\exp ij^a(f_a)$ if $g(x)=\exp
if_a(x)X^a$. 

Similarly, the orientation-preserving diffeomorphisms of $S^1$ are
automorphisms of the CAR algebra via $\psi(f)\mapsto
\psi(\gamma'^{\frac12}f\circ \gamma)$. Again, these are implemented by
unitaries $V(\gamma)$, giving rise to a projective representation of
$\Diff_+(S^1)$ on the Fock space. This representation is related to the
fermionic stress-energy tensor by $V(\gamma)=\exp iT_F(\delta)$
(again, up to a cocycle) if $\delta$ is an infinitesimal
diffeomorphism and $\gamma=\exp(\delta)$.  

On the other hand, for $\gamma\in \Diff_+(S^1)$, the map $g\mapsto
g\circ\gamma$ is an automorphism $\alpha_\gamma$ of the loop group
$LG$. Hence, for any projective representation $\pi$ of $LG$, one
obtains another projective representation $\pi\circ \alpha_\gamma$. 
It turns out that this representation is unitarily equivalent to
$\pi$, i.e., $\alpha_\gamma$ are unitarily implemented by operators $V(\gamma)$ 
giving rise to a projective representation of $\Diff_+(S^1)$
on the representation space of $\pi$. This is the AQFT analogue of the 
Sugawara construction. 

One might object that these theories are just subtheories of free
QFT. However, the Sugawara stress-energy tensor of these theories is different from the
free-field stress-energy tensor, indicating that the dynamics is
different. Accordingly, the current algebras possess many
positive-energy representations that do not arise by restriction of
free-field representations. The study of the representation theory of
chiral CFT models has been in the focus of interest for three decades,
and has revealed a host of fascinating connections, including modular
invariance of characters \cite{V} and $A$-$D$-$E$ classifications \cite{CIZ}.

\subsection{Bosonization}
\label{s:boso}

A much less-to-be-expected ``converse'' of fermionization is
bosonization. It goes back to Mandelstam's vertex operator
construction of free Fermi fields as the exponential of the
(non-existing) massless scalar field, where the latter is to be
written as an integral over the chiral current. Clearly, this integral
makes the construction highly nonlocal. Frenkel and Kac \cite{FK}
rediscovered this (formal) construction in a clean mathematical setup
in terms of well-defined infinite normal ordered exponentials of the
Fourier modes of the current on $S^1$ on the vacuum Fock space of the
current, times a quantum-mechanical factor for the zero mode. The 
latter factor requires to extend the Hilbert space by a factor
$L^2(S^1)$ on which the total charge operator acts like a rotation
with discrete spectrum. The construction therefore selects a discrete
one-dimensional lattice from the continuum of charges (superselection
sectors) of the $u(1)$ current.  

By exponentiating several $u(1)$ currents with coefficients that take
values in an even higher-dimensional lattice, one can construct new
{\em local} fields in a similar way. Some of these theories coincide
with nonabelian current algebras, as in \sref{s:fermi}, at level
$k=1$, where the original $u(1)$ currents are the currents for the
Cartan subalgebra.  

In the AQFT framework, the analogous construction is understood as
a crossed product of the Weyl algebra corresponding to the abelian
current algebra by a lattice subgroup of the continuous group of DHR
automorphisms \cite{BMT,CPS}. This amounts to an extension of the
quasilocal algebra by charged intertwiners (``fields'') whose charges
take value in the lattice. The commutation relations of the fields are
determined by the statistics operators of the DHR automorphisms, which
turn out to be just complex phases. The extension by the new fields is a
local extension if and only if all phases are $=1$, which is precisely
the condition that the lattice is even. 

A lattice extension corresponding to the ``moonshine'' vertex operator
algebra has been constructed using the $24$-dimensional Leech lattice
\cite{KL3}. Its central charge is $c=24$, its $\mu$-index is 1 (i.e., it
has no nontrivial superselection sectors), its vacuum character is the
modular invariant $J$-function and its automorphism group is the
Monster group.  

\subsection{Orbifold constructions}
\label{s:orbi}

One can always descend from a chiral CFT with a compact gauge group of
inner symmetries to the gauge-invariant subtheory. There are many
examples with finite gauge groups; but one may as well consider the
global symmetry of the Lie group $G$ on a current algebra associated
with its Lie algebra $\Lg$. The fixed point subalgebra contains the
Sugawara stress-energy tensor, but is in general larger, with very few
exceptions. 

The formulation as fixed points under the action of a gauge group by
automorphisms of the net of local algebras is the same in the AQFT framework.  

\subsection{Simple current extensions}
\label{s:simpcurr}

The positive-energy representations (sectors) do in general not
form a group under the fusion product; in particular, there is no
inverse but only a conjugate such that the product with the conjugate
contains the trivial (=vacuum) sector. Sectors which have an inverse 
(i.e., the fusion product is the trivial sector), are called simple 
sectors (or ``simple currents'' in some communities). In the DHR
theory, simple sectors are given by DHR automorphisms, rather than 
endomorphisms.

Simple current extensions are extensions of a chiral CFT by local
fields that carry the charge of simple sectors. By the Spin-Statistics
Theorem, such fields can only be local if they carry integer spin,
hence the statistics phases must be $=1$. In order to define the
extensions, one has to specify consistent algebraic relations between
the new fields and the old fields, and among the new fields. The
problem can often be reduced to a control of the representation
category of the extended theory, in terms of that of the original
theory. 

The simple sectors with trivial statistics phases form a group under
the fusion product. In the AQFT framework, simple current extensions
can be defined as crossed products of the original net by the action of
this group, in much the same way as the lattice extension of Weyl algebras in
\sref{s:boso}. 

\subsection{The coset construction}
\label{s:coset}

If a local QFT contains a sub-theory, the ``coset QFT'' is generated
by all local fields of the larger theory that commute with the
sub-theory. Thinking of the latter as generators of a symmetry
(currents as generators of gauge transformations, the SET as generator
of diffeomorphisms), the coset fields are invariant under that
symmetry. Specifically, an inclusion $\Lh\subset \Lg$ of Lie
algebras induces an inclusion of the current algebra chiral CFTs. Both
these CFTs have their own Sugawara stress-energy tensor. Both
stress-energy tensors are generators of the same (universal)
diffeomorphism transformations of the $h$-currents, which means that
their difference commutes with these currents. One obtains the ``coset
stress-energy tensor'' $T_\Lg-T_\Lh$ with central charge
$c_\Lg(k)-c_\Lh(k')$ \cite{GKO}.  

Since the coset
stress-energy tensor is different from the pair of given stress-energy
tensors, the coset theory has its own dynamics. In particular, in this
way all stress-energy tensors with central charge $c<1$ have been
constructed \cite{GKO}, thus rounding off the classification of
admissible values of $c<1$ ``by exclusion of the continuum''
\cite{FQS} by an existence result for the remaining discrete set. 

The coset construction allows to construct ``new models from
old models'', and is by no means restricted to current algebras
associated with Lie subalgebras $\Lh\subset \Lg$. 

In the AQFT framework, the coset construction is given by a
relative commutant of local algebras, namely, if $A(I)\subset B(I)$
are the local algebras of a chiral CFT and a subtheory, then the local
algebras of the coset theory are
\be
C(I):= A(I)'\cap B(I).
\ee

Some of the Virasoro algebra theories with $c<1$ admit extensions by
further local fields, without increasing the central charge. A
complete classification has been obtained by AQFT methods (building on
the earlier classification of modular invariant matrices), see
\sref{s:Qsys}.

\section{Algebraic constructions \rm (not only conformal)}
\setcounter{equation}{0}
\label{s:algebraic}

To construct a model in the AQFT framework, one has to specify the
local algebras along with the automorphic action of the spacetime
symmetries, and one has to provide a vacuum representation with
the appropriate spectral properties.  

Starting from a given model, one possibility is to extend it by
enlarging the local algebras. In order to exclude trivial ``extensions
by tensor products'', one would require the extension to be
irreducible, i.e., to have trivial relative commutant. The extended
model will in general require a larger Hilbert space. 

Another possibility is to deform the local algebras on the same
Hilbert space, while preserving the local commutativity and
covariance. 
For a more extensive review of this approach, see \cite{L15}.

A third idea is ``holographic'' in the sense that the spacetime
association of observables in a given net of local algebras is
radically reorganized, such that, e.g., the quantum observables of a
chiral CFT are re-arranged to become the observables of a two-dimensional model. 

We shall present examples of these new construction ideas in the sequel. 

\subsection{Superselection sectors and symmetry}
\label{s:symm}

Unbroken inner symmetries give rise to superselection sectors.

Let $G$ be a compact global gauge group with an action by
automorphisms $\gamma_g$ ($g\in G$) on a local QFT $B$ such that
$\gamma_g$ preserve the local algebras and commute with the spacetime
covariance. Let furthermore the vacuum state $\omega_B$ of $B$ be
invariant under $\gamma_g$, i.e., the vacuum does not break the
symmetry, so that the gauge transformations are implemented by a
unitary representation of $G$ on the vacuum Hilbert space $H_B$. 

The net of fixed-point subalgebras $A(O)=B(O)^G$ inherits the
local structure and the spacetime covariance, and $H_B$ carries a
reducible positive-energy representation of $A$. It decomposes into
subspaces $H_\pi$ generated from the vacuum vector by elements of $B$
transforming in some representation $\pi$ of $G$, and each subspace
$H_\pi$ is invariant under the invariants $A$. In particular, the
cyclic subspace $H_A$ generated by $A$ from the vacuum vector is a
proper subspace of $H_B$. 

One can reconstruct the full vacuum representation of $B$ 
from the vacuum state  $\omega_A$ of $A$, via the GNS
construction of the state $\omega_B = \omega_A\circ\mu$, 
where $\mu$ is the conditional expectation from $B$ onto $A$ given by
the Haar average over the group action.

A breakthrough result by \cite{DR}, valid in QFT in four-dimensional
spacetime, shows that the scenario just described is the generic
origin of superselection sectors. It relies on the fact that in
spacetime dimension $>3$ (and often also $=3$), the category of DHR
superselection sectors \cite{DHR} is a {\em permutation symmetric} C* tensor
category, and such categories can be identified with the dual of a
compact group.  
It proceeds by reconstructing from the given local net $A$ and its DHR
category a (unique up to isomorphism) universal ``field algebra'' $B$ 
(which may be graded local) together with an action of a compact gauge
group $G$ such that $A=B^G$ is the fixed-point subalgebra, and the DHR
sectors $\rho$ of $A$ are in 1:1 correspondence with the unitary
representations $\pi$ of $G$ with statistical dimension $\dim(\rho)=$
matrix dimension $\dim(\pi)$. 

As a consequence of this result, every irreducible extension of $A$ is
the intermediate algebra of invariants of $B$ under a subgroup $H\subset G$. 

It may appear natural to expect some inner symmetry also to be at the
origin of superselection sectors in low-dimensional spacetime. 
However, there are obstructions due to the fact that the braiding is
not a permutation symmetry in low dimensions (which is in turn a
consequence of the geometric property that the causal complement of a
finite connected region has two connected components). As a
consequence, the dimensions $\dim(\rho)$ fail to be integer in
general, and a 1:1 association with 
finite-dimensional representations $\pi$ of some inner symmetry
such that $\dim(\rho)=\dim(\pi)$ as before cannot be expected. 

Yet, it is possible to associate a (non-unique) weak C* Hopf algebra
\cite{WCH} with the DHR category of a local QFT $A$, where the
non-integrality of dimensions enforces the failure of the property
$\Delta(1)=1\otimes 1$ of the coproduct. The ``reduced field bundle''
$F$ of \cite[II]{FRS} can be interpreted as a nonlocal sector-generating field
algebra with an action of a weak C* Hopf algebra such that the
invariants are the observables $A$. An undesired but unavoidable
feature of this construction is that the embedding $A\subset F$ is not
irreducible, and $F$ contains elements which belong to each of its
local algebras.

\subsection{Extensions by Q-systems}
\label{s:Qsys}

An alternative approach was initiated in \cite{LR95}, by
characterizing irreducible covariant extensions $A\subset 
B$ of a given local QFT $A$. It is assumed that $B$ is relatively
local w.r.t.\ $A$, i.e., observables of $B$ commute with observables
of $A$ at spacelike distance, and that there is a conditional expectation
$\mu:B\to A$ taking $B$ onto $A$ and preserving the vacuum state. 
One does not assume any specific symmetry concept (like group or weak
Hopf algebra), but retains only the conditional expectation as a 
substitute for the Haar average over the action of the gauge group. 

It is also assumed that the index of the local subfactors $A(O)\subset
B(O)$ (which is independent of $O$) is finite, which is automatic if
$A$ is completely rational. 

This scenario includes simple current
extensions as well as orbifold constructions (regarding the full
algebra as an extension of the fixed points), and conformal embeddings
(local CFTs that share the same stress-energy tensor \cite{SW}) as
well as coset constructions (regarding $B$ as an extension of
$A\otimes C$ if $C$ is the coset model of $A\subset B$). But it is more general since it is not required that
the extension $B$ of $A$ is itself local.  

The main result is that every such extension is characterized by a
``Q-system'' in the DHR category (or ``DHR triple''), and every Q-system allows to
reconstruct the extension $B$ in terms of data pertaining solely to
$A$ and its DHR category. 

A Q-system is a triple, consisting of a DHR endomorphism $\theta$ of
$A$ and a pair of intertwiners $w\in\Hom(\id_A,\theta)$,
$x\in\Hom(\theta,\theta^2)$ satisfying the relations 
\be
w^*w= x^*x =d\cdot \eins,\quad w^*x=\theta(w^*)x=\eins,\quad xx =
\theta(x)x,
\ee
where $d^2 = \dim(\theta)$. In a more abstract category setting, a
Q-system is the same thing as a (standard) Frobenius algebra in a C*
tensor category \cite{QSC}, where the category at hand is the DHR
category of superselection sectors. 

From the data of the Q-system, the net $B$ is reconstructed as an 
extension of $A$. It comes equipped with a local structure $A(O)\subset
B(O)$ and covariance, and with a conditional expectation $\mu$
respecting the local structure and commuting with the covariance. The
GNS representation of the state 
\be
\omega_B = \omega_A\circ\mu
\ee
as in \sref{s:symm} gives the vacuum representation of $B$, which --
as a representation of $A$ -- is equivalent to the DHR representation
$\theta$. To every subsector $\rho\prec\theta$ corresponds a generator
$\psi_\rho$ of $B$ that interpolates the vacuum subspace to the
subspace carrying the representation $\rho$, and that implements
$\rho$ ``in the average'', namely 
\be
\rho(a)=\mu(\psi_\rho \, a \, \psi_\rho^*).
\ee
The algebraic relations among the ``charged fields'' $\psi_\rho$  are
encoded in the intertwiners $w$ and $x$ that specify the Q-system.   

The extension is by construction relatively local w.r.t.\ $A$, and it
is local if and only if the intertwiner $x$ satisfies the condition
$\eps_{\theta,\theta}\, x=x$, where $\eps_{\theta,\theta}$ is the statistics operator.   

Thus, exhibiting Q-systems by solving their defining algebraic relations
within the DHR category of the given subtheory, amounts to a
construction of (relatively local or local) extensions. This is a
finite-dimensional problem in rational theories, since for irreducible
extensions, one can show that the multiplicity of every sector
$\rho\prec\theta$ is bounded by its dimension; hence there are only
finitely many ``a priori candidates'' for $\theta$, and the intertwiner spaces,
where $w$ and $x$ take values, are also finite-dimensional. Thus, even
lacking more inspired methods (see below), Q-systems in a given
rational C* tensor category can in  
principle be classified ``by brute force''. 

A very useful fact is that possibly nonlocal chiral extensions
$A\subset B$ induce local two-dimensional extensions $A\otimes
A\subset B_2$, that are ``CFT realizations of modular matrices''. The original construction \cite{R00} uses
``$\alpha$-induction'', and was recently recognized \cite{BKL} to
coincide, in terms of the corresponding Q-systems, with the ``full
centre'', which is a most interesting concept in braided tensor
categories \cite{FFRS06,KR}.  

Complete classifications of local extensions have been achieved \cite{KL1,KL2} for the
chiral and two-dimen\-sional Virasoro nets with $c<1$. The authors have
exploited the fact that the Virasoro nets with $c<1$ are completely
rational, hence their DHR categories are modular (the braiding is
non-degenerate). In this case, one can associate a modular invariant
matrix with every chiral Q-system \cite{BEK99,BEK00}, and these matrices 
have been classified before \cite{CIZ}. Since the DHR representation
$\theta$ of the underlying chiral Q-system can be read off the modular invariant matrix, the number of
candidates for $\theta$ is drastically reduced. The authors then show
existence and uniqueness of the Q-system for each $\theta$ (in the
chiral case) by using more abstract existence and uniqueness results
of \cite{KO}, and they classify the Q-systems for a given $\theta$ by 
a certain second cohomology \cite{IK} in the two-dimensional case.   

It turns out that all local chiral extensions are: an infinite series
of $\ZZ_2$ simple current extensions, and four exceptional
extensions labelled $(A_{10},E_6)$, $(E_6,A_{12})$, $(A_{28},E_8)$,
and $(E_8,A_{30})$ according to the $A$-$D$-$E$ classification
\cite{CIZ} of their modular invariants. Of these, three have been
identified with coset extensions using current algebras \cite{BE}, 
except $(A_{28},E_8)$ which occurs at $c=\frac{144}{145}$.  
This one was later identified with the ``mirror'' of a coset
extension, where the mirror construction \cite{X} is an operation  
on Q-systems relating a Q-system in $A$ to a Q-system in $C$ if $A$
and $C$ are each other's relative commutants (coset models) within
some common extension $B$.

\subsection{Borchers triples and deformation methods}
\label{s:BT}

A net of local algebras $A(O)$ in any dimension can be constructed
from a ``Borchers triple'' (or ``causal triple''). A Borchers triple
consists of a  
von Neumann algebra $M\subset B(H)$ with a cyclic and separating
vector $\Omega\in H$, and a unitary positive-energy representation $U$
of the proper ortho\-chronous Poincar\'e group $\Pc$ on $H$ 
for which $\Omega$ is the unique invariant ground state. It is
required that 
\begin{eqnarray} \label{borch}
U(\lambda)M U(\lambda)^*\subset M &\mathrm{whenever} \quad \lambda
W_0\subset W_0 \notag \\ 
U(\lambda)M U(\lambda)^*\subset M' &\mathrm{whenever} \quad \lambda
W_0\subset W_0',  
\end{eqnarray}
where $\lambda$ stands for $(\ul a,\Lambda)\in \Pc$, $W_0=\{\ul x\in
\RR^{1,D-1}: x^1>\vert x^0\vert\}$ is the standard wedge region of Minkowski
spacetime, and $W_0'=\{\ul x:x^1<-\vert x^0\vert\}$ its causal
complement.  

Clearly, in every QFT, the algebra $M=A(W_0)$ of the standard wedge
in the vacuum representation gives a Borchers triple, by virtue of
covariance and locality. 

Conversely, every Borchers triple defines a net by a simple construction.
The construction proceeds by defining 
$A(W_0):=M$ and $A(\lambda W_0):=U(\lambda)M U(\lambda)^*$ for $\lambda\in
\Pc$. Then one defines
\be
A(O):=\bigcap_{W\supset O} A(W),
\ee
where the intersection runs over all Poincar\'e transforms $W=\lambda
W_0$ of the standard wedge which contain $O$. The assumptions
\eref{borch} ensure that the algebras $A(W)$ are well-defined, and
that the net $A(O)$ is local and covariant. (Unfortunately, the
algebras $A(O)$ may fail to satisfy the Reeh-Schlieder property
($A(O)\Omega$ dense in $H$), and may be as small as $\CC\cdot\eins$.)  

For Borchers triples in two dimensions, the second condition of
\eref{borch} is obsolete (because there are no such $\lambda\in
\Pc$). Moreover, it is sufficient to have a positive-energy representation of the
translations only: one can then use Tomita's Modular Theory to
reconstruct also the representation of the Lorentz group including the
CPT conjugation. Namely, Borchers \cite{Bo} has discovered that the 
inclusions $U(a)MU(a)^*M\subset M$ for $a\in W_0$ are half-sided
modular, and the modular group $\Delta^{it}$ and the modular
conjugation $J$ of $(M,\Omega)$ satisfy the same commutation relations
with the translations $U(a)$ as the Lorentz group $U(\Lambda_{-2\pi
  t})$ and the CPT conjugation $U(\Theta)$, so one can define
$U(\Lambda_t):=\Delta^{-it/2\pi}$ and $U(\Theta):=J$, and the second of
\eref{borch} is automatic for $\lambda$ involving the conjugation. 

(In an effort to generalize this powerful result to $D=4$, K\"ahler and Wiesbrock \cite{KW} have given a
characterisation of the ``relative modular position'' of several von
Neumann algebras with a joint cyclic and separating vector, such that
their modular groups generate the four-dimensional Poincar\'e group.)

Exhibiting Borchers triples amounts to the construction of a
QFT, with any prescribed particle content specified by the representation $U$. 

The difficulty is, of course, to find algebras $M$ satisfying the
assumptions, and to find criteria such that the intersections $A(O)$
are large enough. This is easy for free theories, where $M$ is
generated by the Weyl operators of the free field smeared within
$W_0$. Using Modular Theory, one can also define the wedge algebras
for free fields associated with Wigner's massless ``infinite-spin''
representation \cite{BGL02}, but the last step defining $A(O)$ for
doublecones fails \cite{K15,LMR}: the intersections of algebras turn out to be
trivial unless $O$ contains an (arbitrarily
narrow) infinite spacelike cone. Indeed, ``string-local'' fields
associated with the infinite-spin representation have been constructed
in \cite{MSY}.

The prevailing ideas for finding more examples proceed by deformations
$\wt M$ of free-field (or any other given) algebras $M=A(W_0)$, so
as to produce deformed nets $\wt A(O)$. They are particularly
successful in two dimensions: 

Lechner \cite{L} has constructed integrable massive models with
factorizing scattering matrix by deformations of the canonical
commutation relations (Zamo\-lodchikov-Faddeev algebra). The input in
this approach is a scattering function (in the rapidity variable),
subject to restrictive analyticity conditions. The question whether
the local algebras $A(O)$ are sufficiently large can be answered by a
regularity criterium on the scattering function \cite{L}. In the
affirmative case, the scattering matrix of the resulting deformed
local quantum field theory factorizes into two-particle scattering
matrices given by the input scattering function.

Another deformation method uses ``warped convolutions''
\cite{BLS,A1,A2}. These can be regarded as ``momentum dependent
translations'' of the elements of wedge algebras where the spectrum of the
momentum ensures that wedge-locality is preserved. This deformation
violates Lorentz invariance in more than two dimensions.

Chiral conformal QFT can be used as a starting point to construct
both massive and massless new two-dimensional QFT models 
through ``Longo-Witten endomorphisms'', as follows.

If $A$ is a chiral CFT net, then $(M=A(\RR_+),\Omega,T=U\vert_\RR)$ is
a chiral Borchers triple, namely, the translations $T(a)$ with $a>0$
satisfy $T(a)MT(a)^*\subset M$ for $a>0$. 
In order to get a two-dimensional Borchers triple, one has to
extend $T$ to a unitary representation $T_2$ of the two-dimensional
translations $\RR^2$ with positive energy, such that $T(\ul a)MT(\ul
a)^* \subset M$ for $\ul a\subset W_0$. 

A Longo-Witten endomorphism \cite{LW} is an endomorphism of $M$ of the form
$\Ad_V$, where $V$ is a unitary operator commuting with $T$ and
preserving $\Omega$. Therefore, every one-parameter semigroup of
Longo-Witten endomorphisms $V(b)=e^{-ib\wt P}$ ($b>0$) with
positive generator $\wt P$ gives rise to a two-dimensional 
Borchers triple by putting $T_2(t,x)=T(t+x)V(t-x)$.   

For the net $A$ generated by the free chiral current, semigroups of
Longo-Witten endomorphisms that arise by second quantization of a
unitary semigroup $V_1(b)$ on the one-particle subspace, have been
classified \cite{LW}, namely $V_1(b)$ turn out to be ``singular
symmetric inner functions'' $\varphi_b(P_1)$ of the chiral one-particle 
momentum operator $P_1$. (A symmetric inner function is the boundary
limit of an analytic function on the upper complex halfplane with 
$\vert\varphi(p)\vert=1$ and $\varphi(-p)=\ol{\varphi(p)}$ for
almost all $p\in\RR$. These conditions precisely ensure that the unitary
$V=\varphi(P_1)$ implements a Longo-Witten endomorphism at the one-particle
level. Symmetric inner functions are closely related to the admissible
scattering functions in Lechner's massive deformations \cite{L}, but
the physical significance of this relation remains to be explored.)
To get a one-parameter semigroup, $\varphi_b(p)=e^{ibf(p)}$
must be singular, i.e., it must not have zeros in $\CC_+$.

The corresponding generator $\wt P_1=-f(P_1)$ is positive iff 
$f(p)=-m^2 p\inv$ for some $m^2>0$. The resulting two-dimensional QFT
with chiral translations $P_1$ and $\wt P_1$ is just the free massive
scalar field (defined on the Hilbert space of the free bosonic
current), because $\ul P_1^2 = P^+_1P^-_1 = P_1 \wt P_1 = 
m^2\cdot\eins$. This is the converse of the well-known fact that the
restriction of the massive free field to a light ray is the
conformal free current. 

The interesting challenge is to find other
one-parameter semigroups of Longo-Witten endomorphisms with 
positive generator that are not of this simple (second-quantized) form. 

Pursuing this idea, Tanimoto \cite{T12}
constructed a large class two-dimensional (massless) Borchers
triples from a chiral Borchers triple $(M=A(\RR_+),\Omega,T=U\vert_\RR)$.  
These constructions proceed by deformations of the 
two-dimensional tensor product theory $A\otimes A$ whose Borchers
triple is $(A(W_0) = M\otimes M',\Omega\otimes\Omega,T_2)$ with
$T_2(\ul a)=T(a^+)\otimes T(a^-)$. 
The deformations act by conjugations on the subalgebras
$M\otimes \eins$ and $\eins\otimes M'$ with different unitary
operators:
$$M_V=V(M\otimes\eins)V^* \vee V^*(\eins\otimes M')V.$$
With suitable conditions on the unitary operator $V$, the 
triple $(M_V,\Omega\otimes\Omega,T_2)$ is a deformed
two-dimensional Borchers triple, whose energy-momentum
spectrum is unchanged. With an appropriate adaptation
of scattering theory to the massless situation, the nontrivial
scattering matrix of this deformed QFT coincides with the square
$V^2$ of the deformation unitary. 

Depending on the choice of $V$ as a function of the chiral momentum
operators, one obtains models that are equivalent, respectively
\cite{LST,T12}, to a massless version of the 
integrable deformations by a scattering function as in \cite{L}, or to
the deformations by warped convolutions \cite{BLS}. Yet different
choices of $V$ of the form $e^{i\kappa Q\otimes Q}$, where $Q$ is the
generator of an inner symmetry of the chiral theory and $\kappa$ a
real deformation parameter, give new classes of deformed models \cite{T12,BT}.  
 
Starting instead from a {\em two-dimensional} (massive) Borchers triple
$(M=A(W_0),\Omega,T=U\vert_{\RR^2})$, the tensor
product $(M\otimes M,\Omega\otimes\Omega,T^{(2)})$ with $T^{(2)}(\ul a)= T(\ul
a)\otimes T(\ul a)$ is the Borchers triple of two uncoupled identical
QFT models. A deformation interaction is introduced by deforming the
wedge algebra: 
$$M^{(2)}_V :=V(M\otimes\eins)V^* \vee V^*(\eins\otimes M)V,$$
where $V$ is again of the form $e^{i\kappa Q\otimes Q}$ with a suitable
self-adjoint generator $Q$. Depending on this choice, the
Reeh-Schlieder property of the local algebras can be established \cite{T14}.

\subsection{Holographic models}

Let $B$ be a chiral QFT net on the real line. For any pair of
intervals $K\subset L$ with non-touching end points, let the intervals
$I$ and $J$ be the two connected components of $L\setminus\ol K
= I\cup J$ such that $I>J$ (elementwise). Let  
\be
O=I\times J:=\{(t,x)\in \RR^{1,1}: t+x\in I,t-x\in J\}.
\ee
Then $O$ is a doublecone contained in the Minkowski halfspace
$\RR^{1,1}_+=\{(t,x)\in \RR^{1,1}:x>0\}$. 

Defining
\be\label{holo}
B_+(O) := B(K)'\cap B(L),
\ee
one obtains a local net of local algebras on the halfspace,
covariant under the diagonal of the direct product of two
diffeomorphism groups, acting on $t+ x\in \RR$ and $t-x\in\RR$, respectively. 

This is the prototype of a ``holographic'' construction, since every
local operator of $B$ (a one-dimensional net on $\RR$) is a local
observable of $B_+$ (a two-dimensional net on $\RR^{1,1}_+$), but the
localization assigned to it is very different. In the conformal case,
this is precisely the algebraic AdS-CFT correspondence \cite{AH},
where $\RR^{1,1}_+$ appears as a chart of the two-dimensional Anti-deSitter spacetime. 

In order to ensure locality of $B_+$, 
it is actually not necessary that $B$ is local: it is sufficient that
$B$ is relatively local w.r.t.\ a local subnet $A$ on $\RR$. In this
case, $B_+(O)$ contains at least the subalgebra $A_+(O):=A(I)\vee A(J)$, but
already for $B=A$, $B_+(O)$ is strictly larger than $A_+(O)$
whenever $A$ possesses nontrivial DHR sectors. 

E.g., if $A$ is a Virasoro net, then the halfspace net with local
algebras $A_+(O)=A(I)\vee A(J)$ has the obvious physical interpretation as
the algebra generated by a two-dimensional stress-energy tensor
localized in $O$, whose chiral components $T_+=T_-$ are identified by the
boundary condition $T^{01}(t,x)\vert_{x=0}=0$. This condition is just
the conservation of energy at the boundary $x=0$.  

More generally speaking, the holographic halfspace models \eref{holo}
are extensions of chiral halfspace CFTs with local algebras
$A_+(O)$, which arise by means of a boundary condition
on two-dimensional chiral fields; and every such extension is of the
form \eref{holo} (or intermediate between $A_+(O)$ and $B_+(O)$)
\cite{LR04}.   

There also exist algebraic prescriptions to ``remove the boundary''
\cite{LR09} and to ``add a boundary'' \cite{CKL}, which allow to pass
between extensions $A_+\subset B_+$ on the halfspace and extensions
$A_2\subset B_2$ of CFTs on two-dimensional Minkowski spacetime, where
$A_2(O) = A(I)\otimes A(J)$ is the local algebra of a pair of
independent (although isomorphic) chiral algebras. Under the local
isomorphism (in completely rational models) $A(I)\vee A(J)\sim
A(I)\otimes A(J)$, these pairs of extensions are locally (but of course not
globally) isomorphic.  

As discussed in \sref{s:Qsys}, we can think of $A\subset B$ as a
relatively local chiral extension, described by a Q-system of $A$. Via
the holographic construction, this produces a local extension
$A_+\subset B_+$ on the halfspace, and by ``removing the boundary'',
one arrives at a local two-dimensional extension $A_2\subset B_2$. The
Q-system for $A_2\subset B_2$ as a ``functional'' of the underlying chiral
Q-system for $A\subset B$ turns out to be precisely the
$\alpha$-induction construction \cite{R00} which was discovered
without knowing the steps just described. More recently \cite{BKL}, it
was also identified with the ``full centre'' of the chiral Q-system, as defined
in \cite{FFRS06,KR} in the abstract tensor category setting, providing
yet a different line of construction for the same extension
$A_2\subset B_2$. 

\subsection{Phase boundaries}
\label{s:phase}

In contrast to the ``hard'' boundaries as encountered in the previous 
subsection (physics is defined only in a halfspace), phase boundaries
may separate ``different physics'' on both sides of the boundary. In
two-dimensional conformal QFT, imposing the conservation of energy
{\em and}
momentum at the boundary implies that the two local QFTs on both sides
share the same $2D$ stress-energy tensor, and further boundary conditions
may imply more common chiral observables. 

The issue is thus to have two possibly different local quantum field
theories $B^L$ and $B^R$, each defined in halfspaces $\RR^{1,1}_L$, $\RR^{1,1}_R$,
to be represented on the same Hilbert space such that two requirements
are respected: $B^L$ and $B^R$ share a common subtheory $A$, defined
on full Minkowski spacetime $\RR^{1,1}$, and all observables satisfy
Einstein causality, i.e., any two observables at spacelike
distance commute. 

Because the stress-energy tensor (contained in $A$) generates the full
spacetime covariance, it can be used to extend both halfspace nets to
the full Minkowski spacetime (with their local operators ``on the wrong
side of the boundary'' not being considered as observables). Thus, one
has two full local QFTs with a common subtheory defined on the same
Hilbert space, where Einstein causality for the observables is
equivalent to $B^L$ being ``left local'' w.r.t.\ $B^R$, i.e., a pair
of observables of $B^L$ and $B^R$ commutes if the former is localized
at the spacelike left of the latter. 

This algebraic situation has been analysed in \cite{BPB}, see also
\cite{QSC}. It is found that, if the local extensions $A\subset B^L$ and
$A\subset B^R$ are given by their Q-systems, there is a universal
extension $A\subset C$ given by the ``braided product'' of the
Q-systems. This extensions contains both $B^L$ and $B^R$ as
intermediate extensions, and $B^L$ is left local w.r.t.\ $B^R$. It is
universal in the sense that every irreducible joint representation of 
$B^L$ left local w.r.t.\ $B^R$ is a quotient of $C$. Thus, to classify
such representations, one has to compute the centre of the universal
extension $C$. 

As a linear space, the centre of the algebra $C$ is generated by
neutral products $\Psi_{\rho}^{L*}\Psi_{\rho}^{R}$ of charged fields
from $B^L$ and $B^R$ (where $\rho=\rho_1\otimes\rho_2$ is a DHR sector
of $A\otimes A$ common to both Q-systems). It is more ambitious to
compute the centre as an algebra, in order to determine its minimal
projections. This can be achieved \cite{BPB,QSC} if the underlying
chiral CFT is completely rational and both extensions are given as
full centres ($\alpha$-induction construction from chiral Q-systems,
see \sref{s:Qsys}): In this situation, the minimal central projections
are in 1:1 correspondence with the irreducible bimodules between the
underlying chiral Q-systems. Each minimal projection assigns numerical
values to the operators $\Psi_{\rho}^{L*}\Psi_{\rho}^{R}$, and thereby
specifies the boundary conditions valid among the charged fields. In
some cases, but not always, they become linear relations of the form
$\Psi_\rho^L=\omega \Psi_\rho^R$ with phase factors $\omega$.  

As an example, we give the classification for the two-dimensional
Ising model, which is originally defined as the continuum
limit of a lattice model of two-dimensional Statistical Mechanics at
the critical point, but can (via an Osterwalder-Schrader ``Wick
rotation'') be regarded as a relativistic quantum field theory. Its
chiral net is given by the Virasoro net $A$ with central 
charge $c=\frac12$. The Ising model is then the unique maximal local
two-dimensional extension $B\supset A\otimes A$, which hase two
charged fields $\Psis$ (the ``order parameter'') and $\Psit$ (the
product of two chiral Fermi fields). 

One finds three bimodules, hence three boundary conditions, expressed
in terms of relations between the charged fields: 
\bea (\mathrm{i}) & \Psit^L=\Psit^R,\quad \Psis^L=\Psis^R; \nonumber \\
(\mathrm{ii}) & \Psit^L=\Psit^R,\quad \Psis^L=-\Psis^R; \nonumber \\
(\mathrm{iii}) & \Psit^L=-\Psit^R,\quad \Psis^{L*}\Psis^R=0. \nonumber
\eea
The first case is the trivial boundary, and the second is the ``fermionic''
boundary where the field $\Psis$ changes sign. The third is the
``dual'' boundary, which allows the coexistence of the order and
disorder parameter fields $\sig$ and $\mu$ (in the original Statistical
Mechanics terminology, see \cite{ST,FFRS04}) on either side: these are
the two isomorphic but independent fields $\Psis^R$ and $\Psis^L$.

\section{Final Remarks}
\setcounter{equation}{0}
\label{s:final}

\subsection{What is special about CFT in two dimensions?}
\label{s:2Dspecial}
Quantum field theory in two dimensions offers a wealth of algebraic 
methods for the construction of models, especially conformal models. 
Why are these methods so efficient in two dimensions, and can one put
them into perspective with QFT in four dimensions? 

\smallskip

1. The prominent reason is the kinematical simplicity of CFT in 2D,
especially the existence of chiral fields related to conserved
tensors fields. Chiral fields live in a ``one-dimensional spacetime'',
namely the light ray, which shares crucial features of space {\em and}
time: local commutativity and spectral positivity of the generator of
translations. 

As a consequence, chiral commutators are ultralocal ($\delta$
functions), supported only in coinciding points rather than in or on a
lightcone). Based on this feature, the L\"uscher-Mack theorem
provides an explicit form of the possible commutators of the
stress-energy tensor, with the central charge as the only free parameter.
A similar parametrization of the commutator in higher dimensions, or
without conformal symmetry, is not known.  

Moreover, the algebra of the stress-energy tensor field (and also of
chiral fields of scaling dimension 1) is that of an
infinite-dimensional Lie algebra, permitting the application of
highest-weight representation methods for the efficient study of its
positive-energy representation theory. In contrast, Lie
fields in 4D (whose commutators are linear in the field) don't exist
\cite{B}. (The argument, based on geometric properties of lightcones
and the spectrum condition, was worked out only for scalar fields, but
is presumably true more generally.)  

\smallskip

2. Another consequence of the ultrolocal commutation relations of chiral
free Fermi fields is the feature that chiral gauge transformations
$\psi(x)\mapsto e^{i\alpha(x)}\psi(x)$ are {\em automorphisms} of the
free Fermi (CAR) algebra. Hence, current fields are their
infinitesimal generators, and can be constructed algebraically by
exploiting this property.

In contrast, in $D\geq 2$, the free Fermi algebra is not gauge
invariant, and gauge invariance requires the coupling to a gauge
field. Currents are generators of gauge transformations at a fixed
time only. Notice, however, that in two dimensions the massless free
Dirac field $\psi$ is nothing but a pair of two chiral Fermi fields
$\psi_\pm(t\pm x) = P_\pm\psi(t,x)$, where $P_\pm =
\frac12(1\pm\gamma_0\gamma_1)$. While local gauge transformations of
the general form $\psi(t,x)\mapsto
e^{i(\alpha(t,x)+\beta(t,x)\gamma_0\gamma_1}\psi(t,x)$ do not preserve
the equation of motion, the chiral gauge transformations are of the more  
special form $\psi(t,x)\mapsto
e^{i(\alpha_+(x^+)P^+ + \alpha_-(x^-)P^-}\psi(t,x)$. The latter commute
with $\gamma^0\gamma^\mu\partial_\mu=P_+\partial_-+P_-\partial_+$,
hence preserve the free equation of motion and the commutation
relations at all times. 
In fact, they may be regarded as gauge transformations of the Cauchy
data. 
Thus, the chiral gauge transformations are the subgroup of the gauge
group which is a symmetry without the interaction with a gauge field.  

\smallskip

3. For general conformal fields in two dimensions, the chiral
factorization manifests itself (in rational theories) in
the form of the conformal block decomposition of their correlation
functions \cite{BPZ}. The analytic behaviour of conformal blocks under
field exchange can be formulated algebraically as an ``exchange
algebra'' of chiral components, which are in turn most naturally 
understood in terms of charged intertwiners among the chiral sectors
subject to braid group statistics \cite[II]{FRS}.  

It is worth a remark that also in four dimensions, conformal partial
waves -- which are akin to but not exactly the same as conformal
blocks -- exhibit a factorization into ``chiral'' factors
\cite{DO}. This feature has been exploited for the study of Wightman
positivity (positivity of the Hilbert space inner product defined by
$2n$-point correlation functions) \cite{NRT,NRW}; but the algebraic
counterpart of an underlying exchange algebra, as suggested in \cite{S},
has not been identified.  

\smallskip

4. Many classification results obtained in two-dimensional conformal QFT 
have been obtained thanks to rationality (finitely many
positive-energy representations), or related properties (strong
additivity, split property, finite $\mu$-index) in the AQFT
framework. These results are then owing to the ensuing rigidity
of the DHR category with finitely many irreducible sectors.  

QFT models with finitely many sectors exist also in four dimensions --
e.g., whenever the global gauge group as in \sref{s:symm} 
is a finite group. The corresponding classification results, also in
the case of compact gauge groups, are all in terms of groups and
subgroups, and do not exhibit a comparably rich structure as in the
low-dimensional case with sectors with braid group statistics.  

\smallskip 

5. Recall that every local QFT can be encoded in a Borchers triple
$(M,\Omega,U)$, see \sref{s:BT}. The defining properties of Borchers
triples include the proper adjoint action of the representation
$U$ of the Poincar\'e group on the von Neumann algebra $M$. In four
dimensions, these are quite difficult to satisfy, and compelling ideas
how to construct such triples (if one does not want to {\em start} from a
QFT) are lacking. In contrast, in two dimensions, only a posive-energy
representation of the translation group is required, and the
representation of the Poincar\'e group can be constructed from the
former with Modular Theory (Borchers' Theorem), and its required
properties are automatic. Therefore, Borchers triples are much easier
to obtain in two dimensions, and their data can be subjected to
algebraic deformations while preserving their defining
properties. This appears to be a promising non-perturbative approach
to obtain new QFT models by deformation of given ones. This idea does
not require conformal symmetry.

\subsection{What can we learn for QFT in four dimensions?}
\label{s:4Dlearn}

Algebraic QFT is a powerful approach that conceptually clarifies many
features known to be true in QFT models, or expected to be true in QFT 
in general. Its value is not least that it allows to sharply exhibit
the (in other approaches often tacit) assumptions that are responsible
for these features. At the same time, it allows to investigate the
consequences when some of these assumptions are not fulfilled -- be it
systematically due to the structure of the two-dimensional spacetime
and its symmetry group, or model-dependently due to specific
properties of the dynamics. 

The ``bifurcation'' between four dimensions and two dimensions has
particularly strong consequences in the theory of superselection
sectors, where the braided tensor category is completely degenerate in
four dimensions, and maximally non-degenerate (modular) in large
classes of two-dimensional conformal QFT models. Yet, it is rewarding
to view either extreme in the light of the other, or in the context of
the general structure, as this opens the mind to the many options QFT
has in store that might be missed by model studies. 

Once the underlying abstract structure has been identified (and
separated from the dynamical details of specific models), the road is
open to classifications. Many classifications have been obtained,
mostly of representation theoretic nature. 

Let us return to the kinematic simplicity of QFT in two dimensions,
with the Poincar\'e group being a subgroup of the product of two
translation-dilation groups. Representations of the
translation-dilation group with positive generator of the
translations can be constructed by Modular Theory \cite{MT}. This feature is 
exploited in new algebraic deformation approaches. The passage to four
dimensions is presently not yet very satisfactory; e.g., the ``warped
convolutions'' deformations \cite{BLS} break parts of the Lorentz
symmetry. In contrast, the modular approach indicated in \cite{KW}
seems not very practical, but it points out a direction: The problem
is to control the relations between the many translation-dilation
subgroups that generate the Poincar\'e group. This is reminiscent of the
classification of semisimple Lie algebras by controlling the relations
between the many $su(2)$ subalgebras that generate them. Gaining
experience with two-dimensional models, one may expect progress also
in four dimensions. 

\bigskip

{\bf Acknowledgment.} I thank Yoh Tanimoto and Jakob Yngvason for a
critical reading of the manuscript. Supported by the German Research
Foundation (Deutsche Forschungsgemeinschaft (DFG)) through the
Institutional Strategy of the University of G\"ottingen.  
The hospitality and support of the Erwin Schr\"odinger International
Institute for Mathematical Physics, Vienna, is gratefully
acknowledged.

\end{document}